\documentclass[referee,sn-mathphys-num]{sn-jnl}% Math and Physical Sciences Numbered Reference Style 
\usepackage{graphicx}%
\usepackage{multirow}%
\usepackage{amsmath,amssymb,amsfonts}%
\usepackage{amsthm}%
\usepackage{mathrsfs}%
\usepackage[title]{appendix}%
\usepackage{xcolor}%
\usepackage{textcomp}%
\usepackage{manyfoot}%
\usepackage{booktabs}%
\usepackage{algorithm}%
\usepackage{algorithmicx}%
\usepackage{algpseudocode}%
\usepackage{listings}%
\begin{document}

\title[Article Title]{Deterministic Creation of Identical Monochromatic Quantum Emitters in Hexagonal Boron Nitride}


\author[1]{\fnm{Muchuan} \sur{Hua}}\email{mhua@anl.gov}

\author[2]{\fnm{Wei-Ying} \sur{Chen}}\email{wychen@anl.gov}

\author[1,3]{\fnm{Hanyu} \sur{Hou}}\email{hhanyu@anl.gov}

\author[1]{\fnm{Venkata Surya Chaitanya}\sur{Kolluru}}\email{vkolluru@anl.gov}

\author[1]{\fnm{Maria K. Y.}\sur{Chan}}\email{mchan@anl.gov}

\author[1]{\fnm{HaiHua} \sur{Liu}}\email{haihua.liu@anl.gov}

\author*[1]{\fnm{Thomas E.} \sur{Gage}}\email{tgage@anl.gov}

\author*[3]{\fnm{Jian-Min}\sur{Zuo}}\email{jianzuo@illinois.edu}

\author*[1]{\fnm{Benjamin T.} \sur{Diroll}}\email{bdiroll@anl.gov}

\author*[1]{\fnm{Jianguo} \sur{Wen}}\email{jwen@anl.gov}

\affil[1]{\orgdiv{Center for Nanoscale Materials}, \orgname{Argonne National Laboratory}, \orgaddress{\street{9700 S. Cass Avenue}, \city{Lemont}, \postcode{60439}, \state{Illinois}, \country{the United States}}}

\affil[2]{\orgdiv{Nuclear Science and Engineering}, \orgname{Argonne National Laboratory}, \orgaddress{\street{9700 S. Cass Avenue}, \city{Lemont}, \postcode{60439}, \state{Illinois}, \country{the United States}}}

\affil[3]{\orgdiv{Materials Science \& Engineering}, \orgname{University of Illinois Urbana Champaign}, \orgaddress{\street{1304 W. Green St. MC 246}, \city{Urbana}, \postcode{61801}, \state{Illinois}, \country{the United States}}}


\abstract{Quantum emitters in two-dimensional materials offer significant potential for quantum information science due to their ease of integration with other materials and devices. However, most existing fabrication methods produce emitters with heterogeneous optical properties, limiting their practical applications. Here, the authors report deterministic creation of identical room temperature quantum emitters using masked-carbon-ion implantation on freestanding hexagonal boron nitride flakes. Quantum emitters fabricated using this approach exhibited thermally limited monochromaticity, with an emission center wavelength of 590.7 ± 2.7 nm, a narrow full width half maximum of 7.1 ± 1.7 nm, excellent brightness with an emission rate of 1 MHz, and exceptional stability under ambient conditions. Density functional theory calculations and scanning transmission electron microscopy suggest that these emitters are comprised of boron centered carbon tetramers. Our method provides a reliable platform for exploring the origins of single-photon emission behavior in hexagonal boron nitride, as well as for the industrial-scale production of quantum technology.}

\keywords{Quantum emitter, Single photon emission, Deterministic creation}



\maketitle

\section{Introduction}\label{sec1}

Quantum information science (QIS) is an emerging field at the intersection of quantum physics, computer science, and information theory. It uses the basic rules of quantum mechanics to handle and process information in ways that traditional computing cannot match. There is, however, no dominant platform technology in QIS and many different materials and devices are actively investigated\cite{Yao:2012,vanderSar:2012,Bernien:2013,Waldherr:2014,Hensen:2015,Bradley:2019,Zhou:2024}. Key areas within this field are photon-based quantum information processing, communications, and transduction. Photon-based systems offer several advantages, including the ability to maintain their quantum state for a long time, minimal interference from the environment, and the capacity to transmit quantum information over long distances through optical fibers\cite{Noda:2000,Raussendorf:2003,Raussendorf:2005,Coste2023}. The essential component of these systems is a quantum emitter (QE), capable of producing single photons that are ideally identical and coherent. There are now many sources of single photon emitters based upon a range of materials platforms and spanning the ultraviolet to infrared spectrum\cite{Lohrmann:2015,Zhou:2018,Wang:2018,Liu:2023}. Although many QEs exhibit excellent properties in isolation, their technological utility also requires deterministic localization and reproducibility across ensembles of QEs.

Two-dimensional (2D) materials, such as transition metal dichalcogenides and hexagonal boron nitride (hBN), have been found to be possible hosts for QEs, attracting significant research interest in the last decade\cite{Rasool:2015,Chakraborty:2015,Grosso:2017,Roy:2021,Tan:2022,Lisong:2022,Benedek:2023}. hBN, in particular, has been extensively studied for QE creation as its large electronic band gap and optical phonon energy, in principle, allow it to host QEs at room temperature with perfect monochromaticity\cite{He:2015,Tran:2016,Ziegler:2019,Xu:2021,Shaik:2021,White:2022}. Existing approaches have successfully achieved spatial control in the creation of QEs; however, most of them have limited control of the spectral properties of generated QEs, limiting practical use. A key factor leading to this problem is insufficient understanding of the true atomic configurations of the defects involved in the observed single photon emission behaviors. To address this issue, many efforts has been undertaken by the global research community\cite{Jungwirth:2017,Hayee:2020,Tan:2022,Sarkar:2024}, and recently, both theoretical and experimental works have suggested that carbon related defects are a major source of the visible (from $435\,\text{nm}$ to $700\,\text{nm}$) QEs in hBN\cite{Tawfik:2017,Shevitski:2019,Mendelson:2021,Likejun:2022,Chen:2023}. Therefore, an improved carbon implantation method for hBN with controlled defect creation is valuable to improve the repeatability and quality of the created QEs, but also significantly reduce the difficulty to identify the correlation between the defects atomic structure and observed single photon emission behaviors. In this study, we report a deterministic QE creation process for freestanding hBN flakes through masked-carbon-ion-implantation (MCI) yielding room temperature QEs which satisfies many important properties for ensembles of QEs. The QEs created using this process have remarkable monochromaticity with 2.5 nm standard deviation in zero phonon line (ZPL) wavelengths, and a linewidth of 6.8 nm full-width-half-maximum (FWHM). In addition to known polarization and single-photon behaviors, they are also exceptionally stable (no observable degradation after days with $10^{5}\,\text{W/cm}^{2}$ laser excitation) and bright (1 MHz emission rate). The origins of the QEs was also discussed, where both photoluminescence (PL) and scanning-transmission-electron-microscopy (STEM) based cathodoluminescence (STEM-CL) analysis suggested that the observed QEs could be categorized into two types: single defects and donor-acceptor pairs. First principles density functional theory defect calculations together with Scanning Transmission Electron Microscopy suggest that these QEs consist of four carbon substitutions. 

\section{Results and Discussion}\label{sec2}

\begin{figure}
    \centering
    \includegraphics[width=1\linewidth]{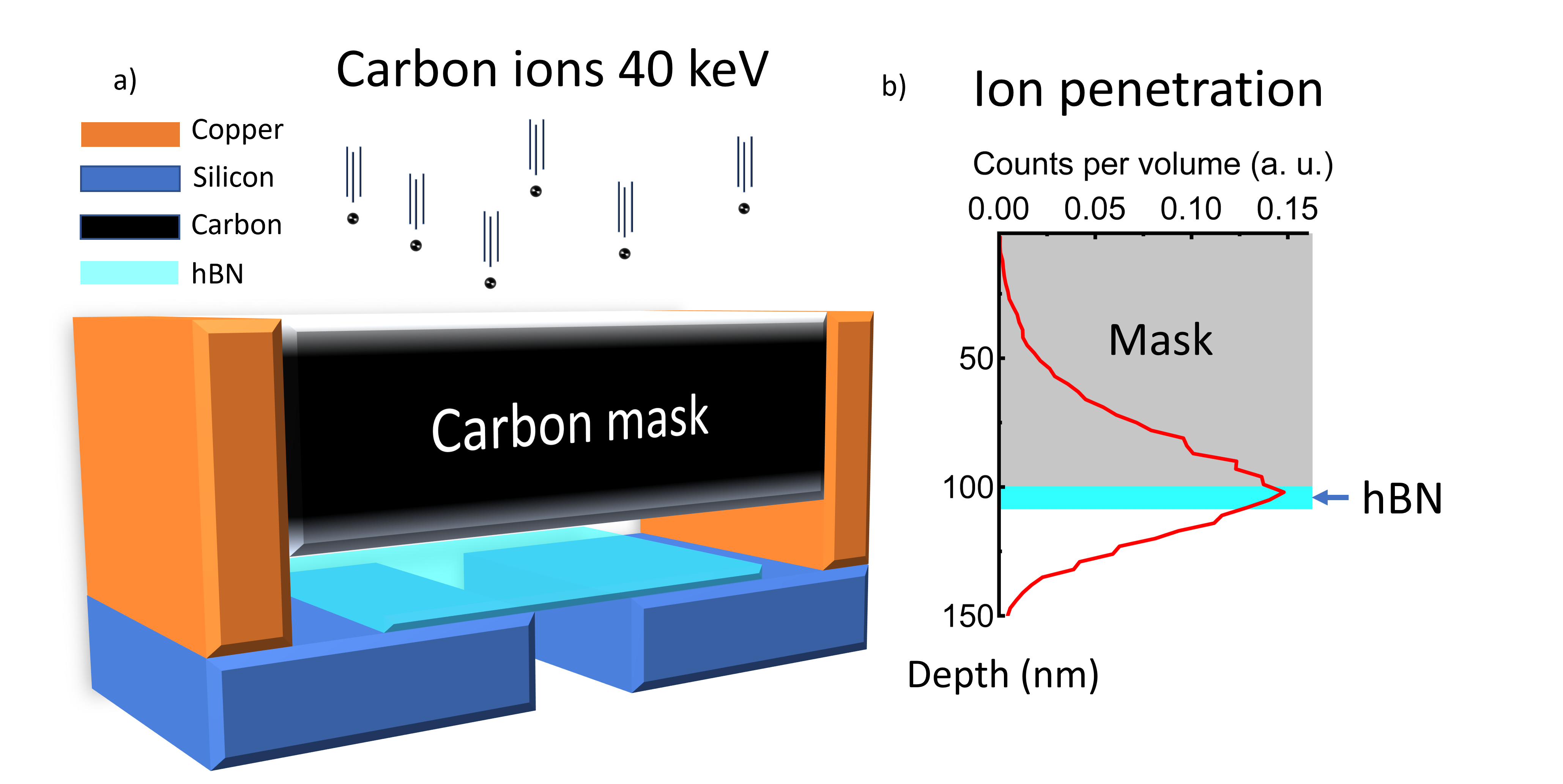}
    \caption{(a) Scheme of the masked-carbon-ion implantation process. (b) Simulated penetration depth of the carbon ions (curve obtained with 100nm carbon mask and 5nm thick hBN flake).}
    \label{fig:sampleconfig}
\end{figure}

\begin{figure}
    \centering
    \includegraphics[width=1\linewidth]{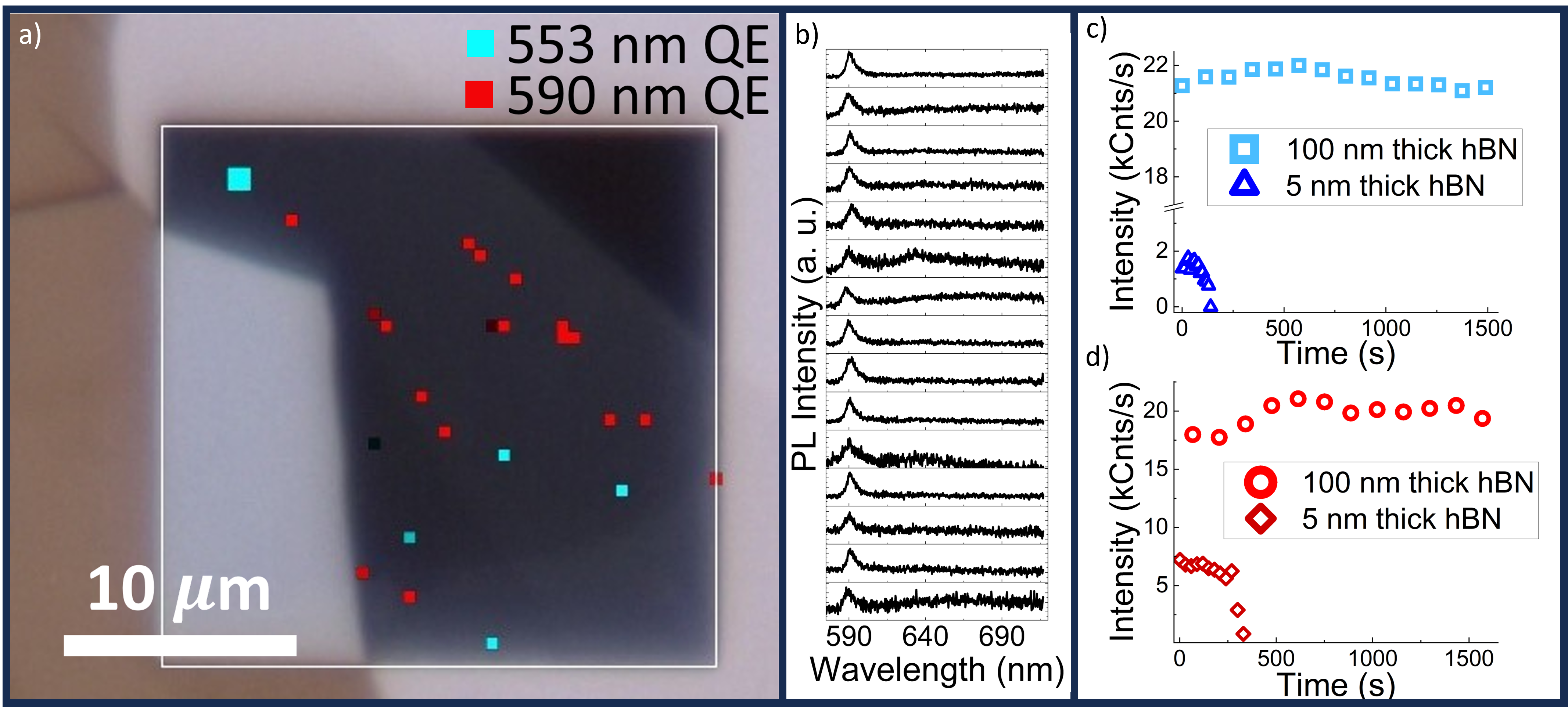}
    \caption{(a) PL hypermap of a typical hBN sample. (b) PL spectra of the 590 nm QEs (marked in red) in the map. (c) Typical PL intensity versus time of 553 nm QEs generated in 5 nm (blue triangles) and 100 nm (light blue squares) thick hBN samples. (d) Typical PL intensity versus time of 590 nm QEs generated in 5 nm (maroon diamonds) and 100 nm (red circles) thick hBN samples.}
    \label{fig:mapping}
\end{figure}

To achieve more consistent creation of carbonaceous defects in hBN with desired ensemble optical properties, typical ion bombardment procedures were modified to minimize lattice damage and generate defects in the middle of the hBN flake. This was achieved by employing an amorphous carbon mask, placed in front of the freestanding hBN flake to decelerate incoming carbon ions (Fig. \ref{fig:sampleconfig}a). The masks were designed and fabricated with a thickness that ensures the carbon ions most probable penetration depth aligns with the middle of the hBN flake, for example, a $100\,\text{nm}$ thick carbon mask was used for a 16 monolayer hBN flake (Fig. \ref{fig:sampleconfig}b) to achieve the maximum stopping efficiency of carbon ions\cite{ZIEGLER20101818}. Such configuration maximized the doping efficiency, leaving less damage due to other defects such as vacancies. Meanwhile, typical contamination caused by the recoiling of the mask elements was automatically precluded as the mask and the implantation ions were made of the same element. To minimize the intrinsic defects of the sample, for instance the defects created during the synthesis, hBN flakes exfoliated from high-quality pristine crystal bulk were used. Since previous work on optical properties of the QEs created by carbon-ion-implantation have found strong substrate dependence\cite{Mendelson:2021}, in this research, the hBN flakes were stamped on to Si or SiN transmission electron microscopy (TEM) grids with pre-fabricated apertures by the dry-transfer method\cite{Castellanos-Gomez:2014} to eliminate the influence from the substrate. As shown in Fig. \ref{fig:mapping}a, a successful sample preparation resulted in a hBN flake with a freestanding area larger than $100\,\mu\text{m}^2$.

During a typical carbon-ion implantation process, a fluence of $2.5\times10^{6}\,\text{ions per}\,\mu\text{m}^2$ with a $40\,\text{keV}$ acceleration energy was applied to achieve a moderate emitter density (around $0.03\,\text{cts per }\mu\text{m}^2$ for sample thickness $<10\,\text{nm}$). Without applying any post treatment, such as high temperature annealing, preliminary scanning of the samples' PL signal were carried out by a reflective confocal microscope system (detailed in methods). Fig. \ref{fig:mapping}a shows the photoluminescence (PL) intensity hyper-mapping of a typical hBN flake treated with MCI, revealing localized QEs under $532\,\text{nm}$ laser excitation. Most observed QEs can be categorized into two types: one with an emission wavelength peaking around 553 nm (colored in cyan in Fig. \ref{fig:mapping}a ) and the other with an emission wavelength peaking around 590 nm (colored in red in Fig. \ref{fig:mapping}a ). For the rest of the paper, these will be denoted as the 553 nm and 590 nm QEs, respectively. The spectra of $590\,\text{nm}$ QEs presented in Fig. \ref{fig:mapping}a are plotted in Fig. \ref{fig:mapping}b, demonstrating extraordinary consistency in their spectral profiles. 
However, the QEs created in samples with thickness less than $10\,\text{nm}$ were generally unstable, with their emission lasting only hundreds of seconds under continuous illumination in ambient conditions. As illustrated in the plots of PL intensity versus illumination time for 553 nm and 590 nm QEs created in a 5 nm thick hBN sample, shown in Fig. 2c (blue triangles) and Fig. 2d (maroon diamonds), respectively, the emission rates diminished after $120\,\text{s}$ and $300\,\text{s}$ respectively. Given the small phonon side band (PSB) observed in these QEs, it is reasonable to infer that the defects are mechanically well isolated, with rigid defect structures and an intact surrounding lattice. Meanwhile, some emitters disappeared after months of storage in air. This is consistent with previous literature which has identified reactions of oxygen with carbonaceous defects in hBN as the source of optical instability in thin flakes (such as 16 monolayers) of hBN\cite{Li:2023}. The thin samples by definition result in implantation of defects close to the hBN surface and also can result in over penetration, which may leave pores for gas transport. Therefore, the authors concluded that exposing QEs to the environment led to their instability.

\begin{figure}[ht!]
    \centering
    \includegraphics[width=1\linewidth]{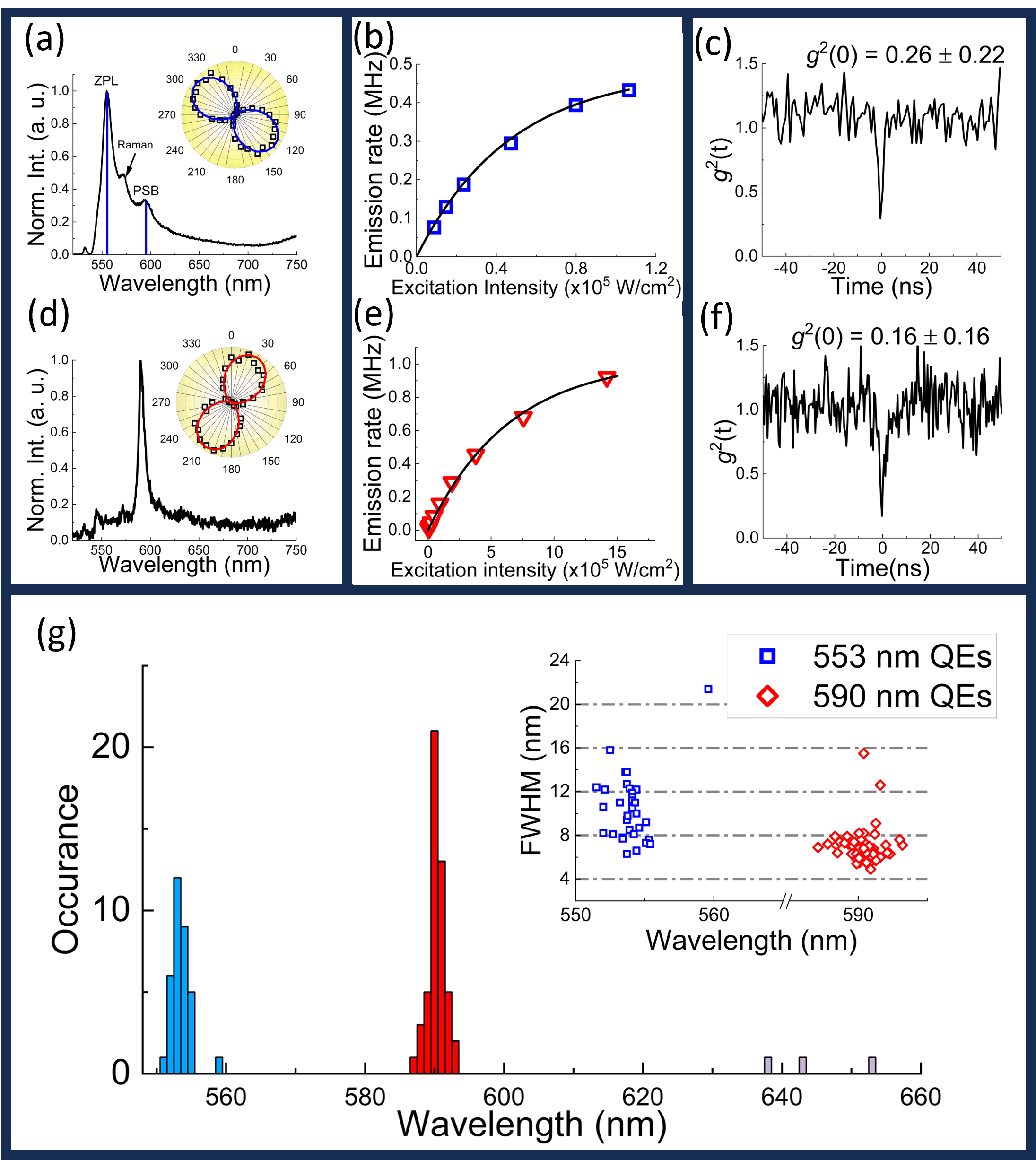}
    \caption{Optical characterization of the QEs. (a) PL spectrum of a typical $553\,\text{nm}$ QE, the insert: PL intensity as a function of the linear polarization angle of the excitation laser. (b PL intensity as a function of the excitation power for a typical $553\,\text{nm}$ QE (c) Photon second-order-correlation function of a typical $553\,\text{nm}$ QE. (d) PL spectrum of a typical $590\,\text{nm}$ QE, the insert: PL intensity as a function of the linear polarization angle of the excitation laser. (e) PL intensity as a function of the excitation power for a typical $590\,\text{nm}$ QE. (f) Photon second-order-correlation function of a typical $590\,\text{nm}$ QE.(g) Emitters' ZPL wavelength distribution with a bar width of $1\,\text{nm}$.The inserts shows the FWHM of the $553\,\text{nm}$ (hollow blue triangles) and $590\,\text{nm}$ (hollow red diamonds) QEs.}
    \label{fig:SPE}
\end{figure}

To solve the stability issue, a $50\,\text{nm}$ carbon mask over a $100\,\text{nm}$ hBN flake sample was selected to achieve a $91.7\%$ stopping efficiency of the carbon ions with negligible over penetration (detailed in supplementary information). After applying MCI to the sample, QEs were created with a total QE areal density of $0.42\,\text{cts per}\,\mu\text{m}^{2}$, an order of magnitude higher than the thin samples with identical ion irradiation fluence. The spectral line profiles of the emitters found in thick films match those of the thin films, but with significantly enhanced stability. Typical PL intensity versus time for 553 nm and 590 nm QEs are shown in Fig. \ref{fig:mapping}c (light blue squares) and \ref{fig:mapping}d (red circles) respectively. No brightness degradation was observed for prolonged illumination with long term fluctuations attributed to the drifting of the stage. The population distribution in wavelength of the QEs created in thin and thick samples are almost identical, where $553.5\pm2.7\,\text{nm}$ and $590.7\pm2.5\,\text{nm}$ (Fig. \ref{fig:SPE}a) were obtained from the combined data. Besides $553\,\text{nm}$ and $590\,\text{nm}$ QEs, other emitters were much less likely ($3.4\,\%$ among the created emitters) to be found in our samples, suggesting the defect creation was dominated by two ultimate processes.

As shown in Fig. \ref{fig:SPE}d, the typical spectrum of a $553\,\text{nm}$ QE was slightly cut on the short wavelength side as multiple $547\,\text{nm}$ long pass filters were used to completely remove the excitation laser light ($532\,\text{nm}$ CW laser). The sharp feature was still pronounced enough for us to estimate the FWHM to be $10.5\pm 3\,\text{nm}$ ($43\,\text{meV}$, blue triangles in the insert in Fig. \ref{fig:SPE}g). Besides the Raman signal of hBN, a secondary peak around $596\,\text{nm}$, $161\,\text{meV}$  from the ZPL, which is slightly lower than the optical phonon energy of hBN, about $170\,\text{meV}$\cite{Reich:2005,Røst:2023}, was observed and has been attributed to the PSB. In contrast, $590\,\text{nm}$ QEs not only showed a much narrower emission profile and smaller dispersion of FWHM, with a value of $\text{FWHM}=7.1\pm 1.7\,\text{nm}$ ($25\,\text{meV}$, red diamonds in the insert in Fig. \ref{fig:SPE}g). A PSB at 636 nm (151 meV from the ZPL) was not pronounced except at high excitation laser power. 

Saturation behavior were observed in both types of QEs with increasing excitation laser power, and the observed maximum emission rates of $553\,\text{nm}$ and $590\,\text{nm}$ QEs were $0.43\,\text{MHz}$ (Fig. \ref{fig:SPE}b) and $0.92\,\text{MHz}$ (Fig. \ref{fig:SPE}e) respectively (considered as lower limits, detailed in Method). Although the maximum emission rate of $553\,\text{nm}$ QEs is much lower than the $590\,\text{nm}$ QEs, they saturate at lower excitation intensity (more than an order of magnitude smaller), suggesting the $533\,\text{nm}$ QEs have a larger absorption cross-section. It is also confirmed in the stability data, where PL intensity's long term fluctuation of the $590\,\text{nm}$ QEs was more pronounced as its collecting efficiency is more sensitive to the drifting the stage due to the small spatial dimension responsible for absorption (confirmed below using CL).

The excitation polarization dependence information of the QEs were obtained by rotating the laser's polarization angle, $\theta$, with a half-wave-plate. The data are shown in the inserts inside Fig. \ref{fig:SPE}d and \ref{fig:SPE}g for $553\,\text{nm}$ and $590\,\text{nm}$ respectively, where the QEs' integrated PL emission intensity $I(\theta)$ can be fitted with the function, 
\begin{equation}
    I(\theta)= I_{0}+I_{\text{A}}*\text{sin}^2(\theta-\theta_{0}),
    \label{eq:polar}
\end{equation}
by assuming the linear dipole response for the defects. Here the purity of the dipoles, $p$ was defined as, $p=\frac{I_{\text{A}}}{I_{\text{A}}+I_{0}}$ with a maximum value of 1, where $I_{0}$ and $I_{\text{A}}$ denote the base value and the amplitude of the fitting function. According to our data, $p$ value approaching $1$ was observed in both types of QEs (typical values of $0.962$ and $0.971$ were observed in $553\,\text{nm}$ and $590\,\text{nm}$ QEs). Such observation indicated an single in-plane transition dipole moment was formed between the defects' ground and excited states.

Use of a Hanbury-Brown-Twiss (HBT) interferometer, the anti-bunching behavior originated from the single photon nature of the defects emission was confirmed(elaborated in the Method section). As shown in Fig. \ref{fig:SPE}f and \ref{fig:SPE}i, $g^{2}(0)$ values of $0.26\pm0.22$ and $0.16\pm0.16$ were obtained for $553\,\text{nm}$ and $590\,\text{nm}$ QEs respectively, confirming their single photon emission behavior with high single photon purity. According to the data, no profound photon bunching at time $t\neq0$ was observed in both types, suggesting the absence of non-radiative shelving state in their electronic states\cite{Michler:2000}. Therefore the QEs' anti-bunching data was fitted by a background corrected two-level-system model\cite{Brouri:2000},
\begin{equation}
    g^{2}(t)= 1- \eta^{2}exp^{-\frac{|t|}{\tau_{\text{r}}}}.
    \label{eq:anti-bunching}
\end{equation}
$\eta=\frac{I_{\text{e}}}{I_{\text{e}}+I_{\text{b}}}$, where $I_{\text{e}}$ and $I_{\text{b}}$ denote the integrated emission intensity and background intensity respectively. $\tau_{r}$ is the defect state's radiative decay mean cycle life time, which is the summation of the radiative excitation and decay lifetimes (statistic of the time-resolved data were obtained by fitting the data with Eq. \ref{eq:anti-bunching}). As shown in Table \ref{tab:QE_info}, short radiative cycle lifetime was observed for both $553\,\text{nm}$ ($1.06\pm0.48\,\text{ns}$) and $590\,\text{nm}$ ($0.69\pm0.17\,\text{ns}$) QEs.  
\begin{table}[]
    \centering
    \begin{tabular}{c|ccccc}
         QE& $\tau_{r}$ & $\eta$  & Peak wavelength & FWHM & Dipole\\
         \hline
         $553\,\text{nm}$ & $1.06\pm0.48\,\text{ns}$ & $0.860\pm0.139$ & $553.5\pm2.7\,\text{nm}$ & $10.5\pm3.0\,\text{nm}$ & Yes\\
         $590\,\text{nm}$ & $0.69\pm0.17\,\text{ns}$ & $0.915\pm0.078$ & $590.1\pm2.5\,\text{nm}$ & $7.1\pm1.7\,\text{nm}$ & Yes
    \end{tabular}
    \caption{Photoluminescence information of the quantum emitters. The values were obtained at room temperature in atmosphere.}
    \label{tab:QE_info}
\end{table}

\begin{figure}
    \centering
    \includegraphics[width=1\linewidth]{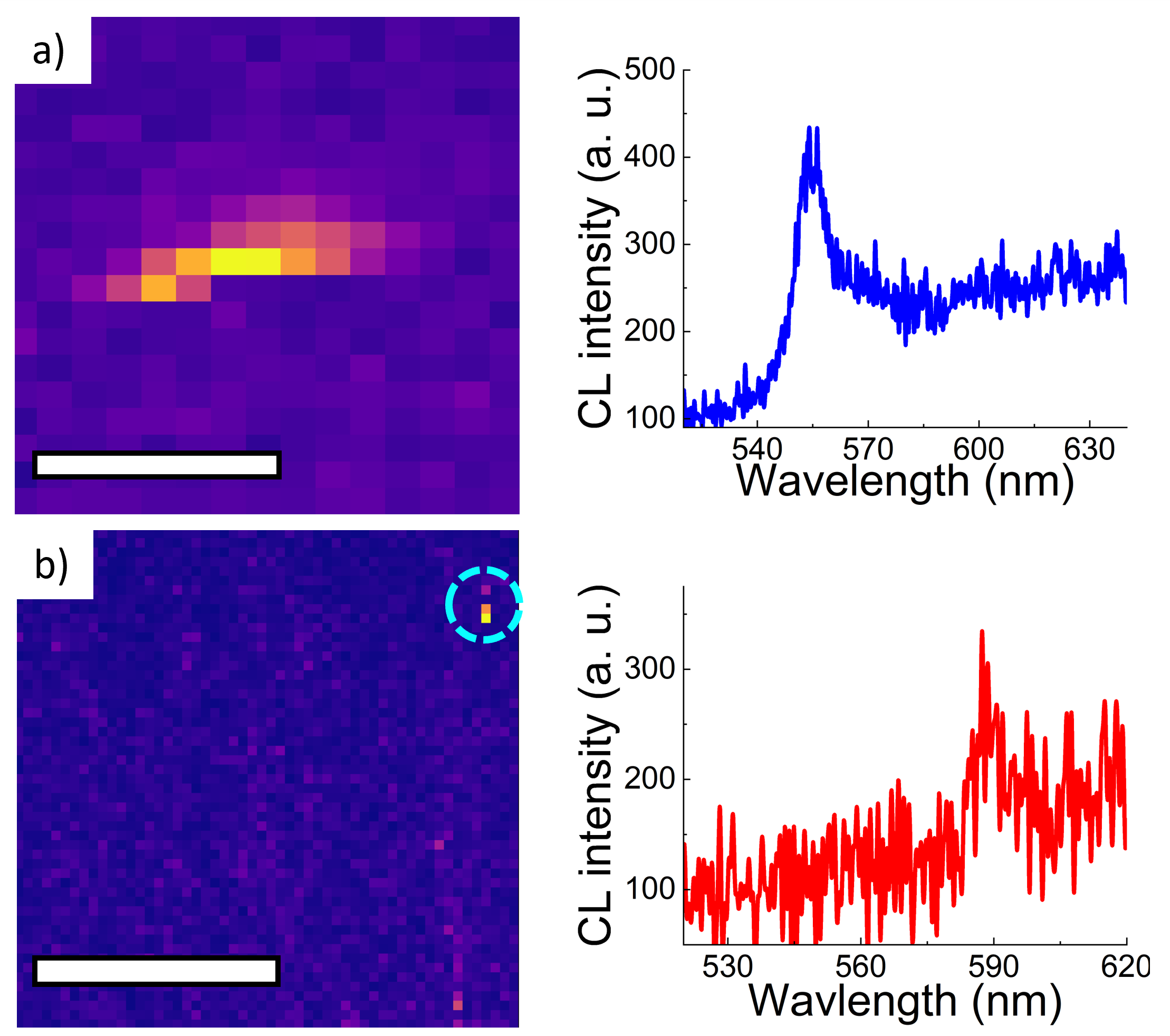}
    \caption{STEM-CL intensity mapping of MCI treated 100nm thick hBN flake. a) Typical mapping of integrated CL intensity (from $547\,\text{nm}$ to $563\,\text{nm}$) with the presence of a $553\,\text{nm}$ QE (the spectrum was derived by the averaging the bright pixels) and its spectrum showing on the right. b) Typical mapping of integrated CL intensity (from $584\,\text{nm}$ to $596\,\text{nm}$) with the presence of a $590\,\text{nm}$ QE (circled with cyan dash lines) and its spectrum showing on the right. The scale bars denote 50 nm.}
    \label{fig:CL}
\end{figure}
The freestanding sample also allows the characterization of the samples with the scanning transmission electron microscope-cathodoluminescence (STEM-CL) system, where the focused electron beam (e-beam) activates the defect related emission. Fig. \ref{fig:CL} shows the CL intensity mapping data collected from an $100\,\text{nm}$ thick sample. The CL spectra was obtained by averaging over the bright pixels within the circled areas. According to the data, the CL spectra of the QEs showed identical wavelength and FWHM to their PL versions. Meanwhile, a much higher spatial resolution than the PL data was achieved, where locations of the 553 nm and 590 nm QEs were determined within 50 nm (Fig. \ref{fig:CL}a) and 4 nm (Fig. \ref{fig:CL}b) respectively. The radius of the excitation volume introduced by the e-beam, $R_e$, can be estimated by the empirical equation: $R_{\text{e}}=(0.0276M/\rho Z^{0.889})E_{\text{b}}^{1.67}$ in the unit of $\mu\text{m}$\cite{Kanaya:1972}. Where $M$ is the molar mass (24.8 g/mol), $\rho$ is the density ($2.1\,\text{g/cm}^3$) and $Z$ is the atomic number (mean value of N and B, 6 was applied) of the target material. $E_{\text{b}}$ denotes the energy of the e-beam in keV. For $80\,\text{keV}$ acceleration energy, which was used in our experiment, $R_{\text{e}}=99.5\,\mu\text{m}$ was estimated, which is much bigger than the thickness of the hBN flakes used in our experiment. Therefore, the excitation volume was roughly defined by the size of the e-beam\cite{Yacobi:1990}, which was much smaller than the size of the observed emission area. So it is reasonable to consider that the interactive volume of the $590\,\text{nm}$ QEs to electron beams was more than an order of magnitude smaller than that of the $553\,\text{nm}$ QEs. This is consistent with the PL measurement, in which photon absorption cross-section of the $590\,\text{nm}$ QEs is smaller than that of the $553\,\text{nm}$ QEs by a factor of $14$, suggesting that the $590\,\text{nm}$ QEs are spatially smaller comparing to the $553\,\text{nm}$ QEs. This difference indicates $553\,\text{nm}$ and $590\,\text{nm}$ QEs originate from different sources, for example, the donor-acceptor-pair\cite{Tan:2022,Singla:2024} and single defects mechanisms respectively. This distinction can adequately explain the observed differences in ZPL linewidth and PSB strength between the two types of QEs. Besides, the CL data also confirm the low emitter density observed in the PL experiment, as no clusters of the emitters were detected, this further highlights the huge difference between the amount of carbon ions stopped by the flake and the resulting emitter density. Here we use the typical emitter density observed in a 100 nm thick hBN flake, $0.42\,\text{cts/}\mu\text{m}^2$ for analysis as they are stable. Compared to the carbon ions stopped by the flake, $2.5\times10^{6}\,\text{per}\,\mu\text{m}^2$, an intuitive speculation is more than one carbon atom was required to form an emissive defect. 

The subsequent analysis only focuses on the 590 nm QEs as they were more likely to have a single defect configuration. The identity of these single defects, however, is not immediately clear. By assuming the stopping of the carbon ions is a completely stochastic process, and the aggregation of carbon ions were negligible, the surface density of $n$ carbon ions replacing neighbouring lattice sites (forming carbon dimers, trimers or tetramers) within a monolayer of hBN, $\Gamma_{\text{n}}$ can be estimated by evaluating the possible configurations of carbon replacement. Since the concentration of carbon ions in each layer was estimated to be low ($<10^{-3}$), for each configuration, a lattice site being intact approaches $1$. Therefore, $\Gamma_{\text{n}}$ can be simply written in the form of (detailed in SI):
\begin{equation}
    \Gamma_{\text{n}}=\Gamma_{s}\cdot\sum_{i=1}^{N}\zeta_{\text{n}}\cdot(\Gamma_{i}\cdot V_{\text{site}})^{n-1}
    \label{eq:density}
\end{equation}
Here $\Gamma_{i}$ denotes the concentration of carbon ions at the $i^{\text{th}}$ layer of the hBN flake, which was obtained from the SRIM (The Stopping and Range of Ions in Matter) simulation. $V_{\text{site}}$ denotes the average volume occupied by each lattice site. $\zeta_{\text{n}}$ denote the number of possible configurations of $n$ carbon ions replacing neighbouring sites. $N$ is the total number of layers of the hBN flake. $\Gamma_{s}$ is the effective surface density measured at the implantation target. $\Gamma_{s}$ and $\Gamma_{i}$ satisfy the relationship: $\Gamma_{s}=\sum_{i=1}^{N}\Gamma_{i}=2.5\times10^{6}\,\text{per}\,\mu\text{m}^2$. According to Eq. \ref{eq:density}, the surface density of carbon dimers and trimers were estimated to be $627$ and $0.185$ per $\mu\text{m}^2$ respectively (detailed in SI). Considering the possible aggregation of the carbon ions inside the hBN flake and the self-repairing processes, the actual density could be higher than the estimation. Suggesting that, most likely, the emissive defects were carbon trimers or tetramers.

\begin{figure}
    \centering
    \includegraphics[width=1\linewidth]{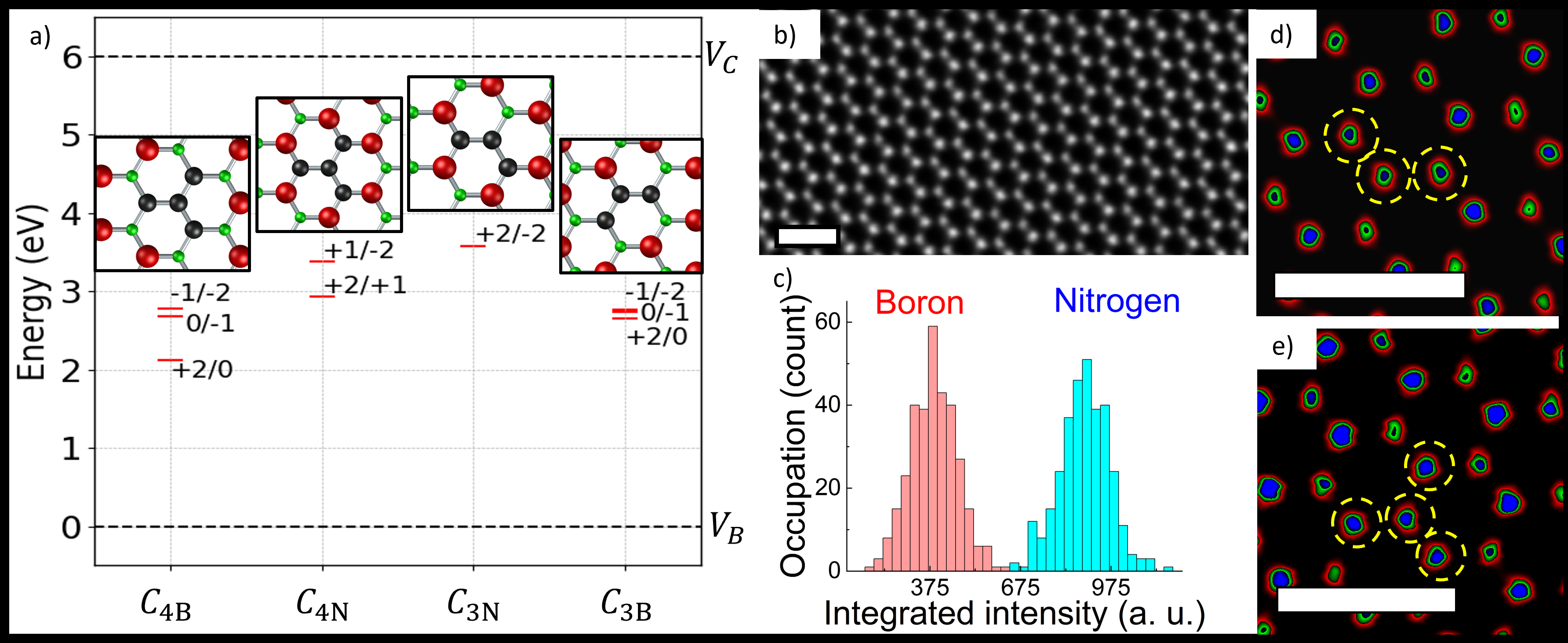}
    \caption{DFT results and the atomic resolution STEM images of the MCI treated hBN sample. a) Allowed electronic energy states derived from DFT calculation for $\text{C}_{4\text{B}}$, $\text{C}_{4\text{N}}$, $\text{C}_{3\text{N}}$ and $\text{C}_{3\text{B}}$ (from left to right). The inserts show the atomic model of the corresponding defects, where green, black and red spheres denote boron, carbon and nitrogen atoms respectively. b) Typical ADF STEM image of the hBN flake. c) Histogram plot of integrated intensity of each dot in b). d) False color STEM image of carbon trimer defects $\text{C}_{3\text{N}}$  replacing "BNB" sites. e) False color STEM image of carbon tetramer $\text{C}_{4\text{B}}$ centered on the boron site. All scale bars denote 500 pm.}
    \label{fig:TEM}
\end{figure}

We utilize first principles density functional theory (DFT) calculations to further validate the proposed defect structures that might be responsible for the observed emissions at 590 nm. We consider the carbon trimer $\text{C}_{3\text{B}}$ (replacing NBN sites, left one insert in Fig. \ref{fig:TEM}a) and $\text{C}_{3\text{N}}$ (replacing BNB sites, left two inset in Fig. \ref{fig:TEM}a) and radial tetramer defects $\text{C}_{4\text{B}}$ (replacing $\text{BN}_{3}$ sites, left three inset in Fig. \ref{fig:TEM}a) and $\text{C}_{4\text{N}}$ (replacing $\text{NB}_{3}$ sites, left four inset in Fig. \ref{fig:TEM}a) and perform DFT calculations with charged defects to predict the defect transition levels within the band gap. Here we applied the charge defect calculations correction methodology \cite{freysoldt2018first} and calculated the charge defect formation energy ($E_{\text{f}}$) as a function of the Fermi energy ($E_\text{Fermi}$), and the results for $\text{C}_{4\text{B}}$, $\text{C}_{4\text{N}}$, $\text{C}_{3\text{N}}$ and $\text{C}_{3\text{B}}$ are shown in Fig. \ref{fig:TEM}a in the order from left to right (details in SI). The calculated electronic states based on the $E_{\text{f}}$ analysis suggests that $\text{C}_{4\text{B}}$ has a transition energy $2.1\,\text{eV}$ corresponding to the $+2/0$ transition, which most closely agrees with the observed emission energy ($2.10\,\text{eV}$).

To corroborate the DFT prediction, atomic resolution STEM images of the samples were taken to reveal the atomic structures involved in the emission. A typical annular dark field (ADF) image of the MCI hBN sample taken over the emitter sites is shown in Fig. \ref{fig:TEM}b, where the hexagonal structure of boron nitride was identified. The image was taken along the direction perpendicular to the flake plane. Here, the hBN crystals used in our experiment showed AA' stacking, where boron and nitrogen sites were alternatively stacking on each other between layers. In such a configuration, two adjacent layers form columns of atoms with identical total $Z$ value, resulting in no contrast difference. Thus, for flakes with an even number of layers, the lattice sites exhibit identical intensity. In contrast, for flakes with an odd number of layers, the intensity difference between boron and nitrogen sites of the non-compensated monolayer is retained. Therefore, Fig. \ref{fig:TEM}b were taken from an odd-layer hBN flake, where the boron (dimmer) and nitrogen (brighter) sites can be well distinguished. A statistical analysis is shown in Fig. \ref{fig:TEM}c, where the histogram of the integrated intensity for each lattice site in Fig. \ref{fig:TEM}b (the image shown was cropped to a quarter of the original one to match the scale) was plotted, showing unambiguous difference between the boron and nitrogen sites. Therefore, lattice sites with ADF image intensity in between the nitrogen and boron sites were attributed to carbon replacements ($Z=6$). As shown in Fig. \ref{fig:TEM}d and \ref{fig:TEM}e, where carbon trimer, replacing BNB sites and carbon tetramer replacing $\text{BN}_{3}$ sites (marked with yellow dashed circles) were observed respectively. The data suggested that $\text{C}_{4\text{B}}$ did form during the MCI process, supporting the emitter density analysis and DFT calculations. 

\section{Conclusion}\label{sec3}
MCI enables repeatable deterministic production of bright, stable and monochromatic room temperature QEs in hBN, ensuring a highly consistent yield of QEs with excellent single photon purity. The mask approach is also inherently amenable to lithography, meaning it is consistent with the ability to place QEs deterministically in space. Our research is both technological and scientific interest, as it provide consistent reproducible QE sample production, allowing reliable platform for characterization and fabrication research. According to our data, boron centered carbon teramers, $\text{C}_{4\text{B}}$ were suggested to be the origins of the 590 nm QEs, while the 553 nm QEs are attributed to the donor-acceptor pairs induced by carbon-defect clusters. The QEs produced with this method exhibiting exceptional stability, underlining the robustness of the proposed approach. Such reliability is paramount when considering the demands of any practical applications especially the industrial-scale production, where reproducibility and consistency are central to achieving standardized outputs, potentially introducing a transformative dimension to the landscape of quantum technology. 

\section{Methods}\label{sec4}
\subsection{Exfoliation and stamp transfer of hBN\cite{Castellanos-Gomez:2014}}
The hBN pristine crystal brought from \textit{hq graphene} were used in our experiment. The bulk crystals were first exfoliated with \textit{SPV-224PR-MJ} tape brought from \textit{Nitto Denko Cooperation}. Then the exfoliated hBN flakes were transferred onto the polydimethylsiloxane (PDMS) film brought from \textit{Gel-Pak} (PF-40X40-170-X4). Next, the hBN flakes on the PDMS film were stamped onto the Si or SiN TEM grids with pre-milled apertures. Three types of grids were purchased from \textit{Simpore}, Si grids with single aperture window, Si grids with 25 nm thick membrane windows and SiN grids with 50 nm thick membrane window. On membrane grids, the apertures were home-milled and detailed in next section. During the stamping process, heating is provided to reduce the adhesive strength of the PDMS film, where 60 Celsius degree was maintained.

\subsection{Open apertures on TEM grids}
The apertures on the membrane TEM grids were milled with the \textit{Carl Zeiss NVision 40} focused ion beam (FIB) system. The focused gallium ion beam with an acceleration voltage of $30\,\text{kV}$ and a current of $300\,\text{pA}$ was used during the milling process. Where $6\,\mu\text{m}\times6\,\mu\text{m}$ square apertures were made to form a $3\times3$ aperture array on the membrane of the grid. 

\subsection{Carbon deceleration mask preparation}
The 400 Mesh TEM copper grids with carbon film were used as the substrate to grow the carbon deceleration mask. For specific carbon mask thickness, carbon deposition through chemical vapor deposition were used with a typical thickness variation of $\pm10\,\text{nm}$.

For thick samples, carbon film supported on single holed TEM grids (\textit{CFGA1500-Cu-ET}) purchased from \textit{Electron Microscope Sciences}) were used. The grids have an aperture with diameter of $1500\,\mu\text{m}$ and the claimed thickness of 25-30 nm (One mask was tested by the EELS measurement in TEM, resulting in a thickness of $26\,\text{nm}$, confirming the claimed value). In a typical MIC treatment, two grids were stacked together to obtain an effective thickness of 50 nm.

\subsection{Masked-carbon-ion-implantation (MCI)}
The masked-carbon-ion-implantation was done with the ion accelerating source of Nuclear Science and Engineering division of Argonne National Laboratory. The carbon ion was obtained by ionization of carbon oxide molecules. The ions were separated by the acceleration field, where an acceleration voltage 40 kV were applied. Then the carbon ions were focused and guided on the sample grids by magnetic lenses. During the ion implantation, the carbon mask were placed in front of the sample grid to achieve desired ion penetration depth in the hBN flake. The typical fluences of the carbon ion applied in the treatment were: $2.5\times10^{14}$.

\subsection{Preliminary PL mapping}
The preliminary PL mapping was done with the \textit{Renishaw Invia Qontor} confocal Raman Microscope. A \textit{Leica} N PLAN EPI 100x/0.85 BD objective was used to focus the laser onto and collect the PL light from the sample. A 532nm solid state laser was used as the excitation source, where a grating with 600 lines/mm was used providing a spectral resolution around 0.17 nm (typically 542 to 702 nm spectral range was selected). During a typical mapping, step size of $0.3\,\mu\text{m}$ was applied.

\subsection{Confocal spectroscopy with anti-bunching measurement}
The confocal system was built on an \textit{Olympus} IX71 inverse microscope. An \textit{Olympus} MPlan APO 100x/0.90NA BD objective was installed to send laser onto and collect PL light from the sample. A 532 nm diode laser (DJ532-10) purchased from \textit{Thorlabs} was used as the excitation source. Di03-R532-25x36 dichroic beam splitter followed by a BLP01-532R-25 long pass filter purchased from \textit{Semrock} and a FELH0550 long pass filter purchased from \textit{Thorlabs} were applied to separate the laser and the PL light. A retractable 80:20 (reflection:transmission) beam splitter was used to pitch the filtered PL light into a SP2300 spectrometer equipped with 150 lines per mm grating blazed at 500nm purchased from \textit{Teledyne Princeton Instrument}, providing a spectral resolution of $1.48\,\text{nm}$, to search and locate the QEs. Once a QE is located, the beam splitter was retracted, and the PL light was coupled to the inlet of a multi-mod 50:50 fiber beam splitter with $50\,\mu\text{m}$ core and FC/APC connectors (customized with \textit{Thorlabs}). The outlets of the fiber beam splitter were coupled to two \textit{PicoQuant} PDM single photon avalanche diodes (SPAD). Between the each pair of the fiber connector and SPAD, a bandpass (FBH550-40 and FBH590-10 from \textit{Thorlabs} for $553\,\text{nm}$ and $590\,\text{nm}$ QEs respectively) filter was applied to remove "afterglow" of the SPAD chip. A MultiHarp 150 from \textit{PicoQuant} was used as the time-correlated-single-photon-counting (TCSPC) device. The software QuCoa from \textit{PicoQuant} was used to perform anti-bunching measurement and the data analysis.

To perform polarization dependence measurement, a Glan-Laser Calcite polarizer (GL10-A) was first introduced to clean up the polarization of the laser, followed by an achromatic half-wave plate (AHWP10M-580) to select the polarization of the laser. Both the polarizer and the half-wave-plate were purchased from \textit{Thorlabs}. In the collection path, a polarizer was applied to check the polarization response of the emission light.

\subsection{STEM-CL imaging and analysis}
The atomic resolution STEM images were collected by the \textit{ThermoFisher} Spectra-200 probe corrected STEM. Typically, atomic resolution STEM images were taken under an electron acceleration voltage of $200\,\text{kV}$, with the gun lens size of 4 and the spot size of 7. The ADF images were captured by the DF-S camera array. The STEM-CL was performed on the QUEEN-M system, where a \textit{ThermoFisher} Spectra-300 probe corrected STEM was used as the platform to accommodate the \textit{Attolight} M\"{o}nch CL module. The \textit{Mel-built} CL holder was used and the microscope was configured with a gun lens size of 1 and a spot size of 3, leading to a typical e-beam current of $0.360\,\text{nA}$. An integration time of 200 ms was applied for each pixel for the CL hyper-mapping.

\subsection{DFT simulation}
All the density functional theory (DFT) calculations are performed with PBE exchange-correlation functional \cite{perdew1996generalized} with GGA approximation using the Vienna Ab-initio Simulation Package (VASP) code \cite{kresse1996efficiency,blochl1994projector,kresse1999ultrasoft}. We use the nonlocal vdW-DF functional \cite{dion2004van} to account for the van der Waals forces within the 2D layers. We choose the MaterialsProject \cite{jain2013commentary,jain2011high} recommended pseudopotentials within the projector augmented wave (PAW) formalism for B, C, N with 3, 4 and 5 valence electrons respectively. A cutoff energy of 500 eV is used for the plane wave basis, along with an automatic mesh with 50 k-points per inverse angstrom for all the calculations. A sufficiently large vacuum of at least 20 Å is used to separate the periodic images of the hBN layers along the vertical direction. We use Gaussian smearing for all the calculations with a smearing width of 0.05. For charge defect calculations, we perform DFT calculations by adding and removing 1 or 2 electrons from the neutral system. That is, each defect structure is evaluated with a charge state of +2, +1, 0, -1 and -2. A large vacuum of at least 30 Å is used for all the charge defect DFT calculations to account for the unwanted interaction between periodic images of the 2D material in the z-direction. Correction for spurious periodic charge interactions are applied according to the scheme of \cite{freysoldt2018first}. 

\bmhead{Acknowledgements}

Work performed at the Center for Nanoscale Materials, a U.S. Department of Energy Office of Science User Facility, was supported by the U.S. DOE, Office of Basic Energy Sciences, under Contract No. DE-AC02-06CH11357. This research is also supported by QIS research funding from the U.S. Department of Energy, Office of Science User Facility and the Development Fund of Argonne National Laboratory and by Laboratory Directed Research and Development (LDRD) program from Argonne National Laboratory, provided by the Director, Office of Science, of the U.S. Department of Energy under Contract No. DE-AC02-06CH11357. M.K.Y.C. and V.S.C.K acknowledge the support from the BES SUFD Early Career award. Authors would like to thank Dr. David J. Gosztola for his advising on the construction of the confocal microscope system; Mr. Peter Baldo and Mr. Dzmitry Harbaruk for operating the ion accelerator in IVEM-Tandem facility in ANL.

\bmhead{Conflict of Interest}
The authors have no conflicts of interest.

\begin{appendices}




\end{appendices}


\end{document}